\documentclass[twocolumn]{aastex63}

\usepackage{xcolor}

\newcommand{\rr}[1]{{\color{black}#1}}

\begin{document}

\title{The role of natal kicks in forming asymmetric compact binary mergers}
\author{Madeline Oh}
\affiliation{New Trier High School, 385 Winnetka Avenue, Winnetka, IL 60093, USA}
\affiliation{Center for Interdisciplinary Exploration and Research in Astrophysics (CIERA), 1800 Sherman, Evanston, IL 60201, USA}
\author[0000-0002-1980-5293]{Maya Fishbach}
\affiliation{Canadian Institute for Theoretical Astrophysics, David A. Dunlap Department of
Astronomy and Astrophysics, and Department of Physics, 60 St George St, University of Toronto, Toronto, ON M5S 3H8, Canada}
\affiliation{Center for Interdisciplinary Exploration and Research in Astrophysics (CIERA), 1800 Sherman, Evanston, IL 60201, USA}
\author[0000-0001-9879-6884]{Chase Kimball}
\affiliation{Department of Physics and Astronomy, Northwestern University, 2145 Sheridan Road, Evanston, IL 60208, USA}
\affiliation{Center for Interdisciplinary Exploration and Research in Astrophysics (CIERA), 1800 Sherman, Evanston, IL 60201, USA}
\author[0000-0001-9236-5469]{Vicky Kalogera}
\affiliation{Department of Physics and Astronomy, Northwestern University, 2145 Sheridan Road, Evanston, IL 60208, USA}
\affiliation{Center for Interdisciplinary Exploration and Research in Astrophysics (CIERA), 1800 Sherman, Evanston, IL 60201, USA}
\author{Christine Ye}
\affiliation{Stanford University, 450 Jane Stanford Way, Stanford, CA 94305, USA}

\begin{abstract}

In their most recent observing run, the LIGO-Virgo-KAGRA~(LVK)~Collaboration observed gravitational waves~(GWs) from compact binary mergers with highly asymmetric mass ratios, including both binary black holes~(BBHs) and neutron star--black holes~(NSBHs). It appears that NSBHs with mass ratios $q\simeq0.2$ are more common than equally asymmetric BBHs, but the reason for this remains unclear. We use the binary population synthesis code~\textsc{cosmic} to investigate the evolutionary pathways leading to the formation and merger of asymmetric compact binaries. We find that within the context of isolated binary stellar evolution, most asymmetric mergers start off as asymmetric stellar binaries. Because of the initial asymmetry, these systems tend to first undergo a dynamically unstable mass transfer phase. However, after the first star collapses to a compact object, the mass ratio is close to unity and the second phase of mass transfer is usually stable. According to our simulations, this stable mass transfer fails to shrink the orbit enough on its own for the system to merge. Instead, the natal kick received by the second-born compact object during its collapse is key in determining how many of these systems can merge. For the most asymmetric systems with mass ratios $q\leq0.1$, the merging systems in our models receive an average kick magnitude of 255~km~s$^{-1}$ during the second collapse, while the average kick for non-merging systems is 59~km~s$^{-1}$. Because lower mass compact objects, like NSs, are expected to receive larger natal kicks than higher mass BHs, this may explain why asymmetric NSBH systems merge more frequently than asymmetric BBH systems.

\end{abstract}

\section{Introduction}
In the latest observing run of the gravitational-wave (GW) detector network consisting of the Laser Interferometer Gravitational-Wave Observatory (LIGO;~\citealp{2015CQGra..32g4001L}), Virgo~\citep{2015CQGra..32b4001A}, and the Kamioka Gravitational Wave Detector (KAGRA;~\citealp{2021PTEP.2021eA101A}), the LIGO-Virgo-KAGRA Collaboration (LVK) observed confidently \emph{asymmetric} compact binary mergers for the first time -- binaries in which one component is distinctly heavier than the other component. 
The first observation of GWs from a confidently asymmetric binary was GW190412~\citep{2020PhRvD.102d3015A}, a merger between a $\simeq 30\,M_\odot$ black hole (BH) and an $\simeq 8\,M_\odot$ BH with a mass ratio $q \lesssim 0.3$; for GW events $q \leq 1$ by definition. This was followed by the discovery of an even more asymmetric system: GW190814~\citep{2020ApJ...896L..44A}, a merger involving a $\simeq23\,M_\odot$ BH and a $2.6\,M_\odot$ compact object. This $2.6\,M_\odot$ object is either the lightest BH or heaviest neutron star (NS) known. Later in the third observing run, the LVK confidently discovered neutron star-- black hole (NSBH) systems for the first time, each consisting of an NS with mass $\lesssim2\,M_\odot$ and a BH with mass between $5$--$10\,M_\odot$, for mass ratios around $q \simeq 0.2$--$0.3$~\citep{2021ApJ...915L...5A}. 

GW observations suggest that such asymmetric NSBH systems merge at a higher rate than \rr{binary black hole (BBH)} systems with similar mass ratios. \citet{2022ApJ...931..108F} recently analyzed the full population of merging compact object binaries, and found that BBH mergers in which both component masses are bigger than $5\,M_\odot$ have a significantly stronger preference for equal mass ratios (with the pairing probability scaling roughly as $q^{3.5}$) compared to merging binaries in which at least one component is smaller than $5\,M_\odot$ (for which the pairing probability scales roughly as $q^{0.8}$; larger powers of $q$ correspond to a stronger preference for symmetric mergers). Among all masses, the pairing probability increases as the mass ratio approaches unity, implying that symmetric mergers are preferred over asymmetric mergers~\citep{2020ApJ...891L..27F,2022ApJ...931..108F,2021arXiv211103634T,2022ApJ...933L..14L}. 

The observed trends in the mass ratios of merging BBH and NSBH are broadly consistent with expectations from theoretical models.
Many scenarios for the formation of merging BBH and NSBH systems have been proposed, including isolated binary evolution in the galactic field, evolution in young star clusters, stellar triples, and dynamical assembly in dense star clusters or the disks of active galactic nuclei (see~\citealp{2022PhR...955....1M,2021hgwa.bookE..16M} for recent reviews). 
For mergers involving NSs, isolated binary evolution is thought to be the dominant scenario (\citealp{2014MNRAS.440.2714B,2014MNRAS.441.3703Z,2018MNRAS.480.4955F,2018MNRAS.479.4391M,2020ApJ...903....8H,2020ApJ...888L..10Y}; but see also \citealp{2019MNRAS.486.4443F,2020MNRAS.497.1563R,2020MNRAS.498.4088M,2021ApJ...908L..38A}).
According to most models of binary stellar evolution, the resulting BBH mergers tend to be symmetric, with a mass ratio distribution that peaks close to $q = 1$~\citep{2012ApJ...759...52D,2018MNRAS.474.2959G,2019MNRAS.490.3740N}, while predicted NSBH mergers have typical NS masses of $\sim1.4\,M_\odot$ and BH masses of $\sim5$--$10\,M_\odot$, implying mass ratios $q < 0.3$~\citep[e.g.][]{1993MNRAS.260..675T,2002ApJ...572..407B,2018MNRAS.481.1908K,2021MNRAS.508.5028B}.

These predicted NSBH masses depend on the uncertain existence of a ``mass gap" between the most massive NS ($\simeq2.2\,M_\odot$ based on inference of the NS equation of state;~\citealp{2021PhRvD.104f3003L}) and the least massive BH, which X-ray binary observations suggested to be $\simeq 5\,M_\odot$~\citep{2010ApJ...725.1918O,2011ApJ...741..103F}. 
Theoretically, a mass gap may apply to the birth masses of NSs and BHs (e.g. through the core-collapse supernova mechanism; ~\citealp{2012ApJ...749...91F,2012ApJ...757...91B,2020ApJ...891..141G,2022ApJ...931...94F}), or a mass gap may result from the evolutionary sequences that lead to systems like GW sources or X-ray binaries (e.g. natal kicks or mass transfer;~\citealp{2021MNRAS.500.1380M,2022arXiv220906844S,2022ApJ...940..184V}).
If a mass gap exists, all NSBH systems would be asymmetric.
However, irrespective of the mass gap question, asymmetry seems to be more common among GW sources with a low mass component~\citep{2022ApJ...931..108F}. GW190814's $2.6\,M_\odot$ component may be in the purported mass gap, but its extreme mass ratio of $q \simeq 0.1$, rather than its low secondary mass, makes it an outlier from the rest of the BBH population~\citep{2021ApJ...913L...7A,2022ApJ...926...34E}.

Previous studies have explored pathways within isolated binary evolution that lead to asymmetric BBH or NSBH mergers, usually focusing on one type of binary, either BBH or NSBH, and often within the context of the NS/BH mass gap~\citep[e.g.][]{2020ApJ...901L..39O,2020ApJ...899L...1Z,2021MNRAS.508.5028B,2022A&A...657L...6A,2022ApJ...933...86Z}.
In this work, we consider both BBH and NSBH asymmetric systems, and investigate the evolutionary pathways that lead them to form and merge within the context of isolated binary evolution. We treat the issue of asymmetry separately from the NS/BH mass gap, and assume that NSs and BHs are born without a mass gap (the delayed core-collapse supernova prescription from~\citealp{2012ApJ...749...91F}). We focus on how a binary's mass ratio affects and is affected by evolutionary sequences like mass transfer and natal kicks. 

Understanding the interplay between the initial binary mass ratio, its evolutionary sequence, and the final mass ratio requires simulating large populations of binaries and tracking their detailed progressions. We use the \textsc{cosmic} rapid binary population synthesis code to simulate the evolution of large binary star populations under different physical assumptions~\citep{2020ApJ...898...71B,2021ascl.soft08022B}. \textsc{cosmic} simulates binary stars starting with an initial population at Zero Age Main Sequence (ZAMS) and tracking how they evolve in a Hubble time. We compare the evolutionary pathways leading to asymmetric NSBH mergers versus BBH mergers, aiming to to understand (a) the challenges to forming asymmetric mergers in isolated binary evolution (b) the evolutionary sequences that lead to the most asymmetric mergers and (c) the key ingredients to achieving asymmetric mergers, including the role of natal kicks. 

The rest of this paper is structured as follows. In Section~\ref{sec:methods}, we provide an overview of some key stages in isolated binary evolution and describe our simulation setup. Section~\ref{sec:results} details the main results, including a description of the categories into which we sort the evolutionary pathways, an explanation of how asymmetric systems form according to our simulations, and the impact of natal kicks. In Section~\ref{sec:discussion}, we describe the implications of our results as well as the limitations of this work and plans for the future. We conclude in Section~\ref{sec:conclusion}.

\section{Simulations}
\label{sec:methods}
We briefly summarize some key evolutionary stages and their connection to the binary mass ratio in~Section~\ref{subsec:keystages}. We then describe our \textsc{cosmic} simulation settings in Section~\ref{subsec:cosmic}. 

\subsection{Key stages in binary evolution}
\label{subsec:keystages}
Isolated binary evolution generally includes two phases of mass transfer: once when the initially more massive object fills its Roche lobe, and again when its slower evolving companion fills its Roche lobe. 
Different pathways within binary stellar evolution generally vary in the type (stable or unstable) and timing of mass transfer between the two components.
Roche lobe overflow can lead to stable mass transfer, in which the donor star's Roche lobe expands faster than the stellar radius and the mass transfer shuts off. In other cases, the mass transfer may be unstable, so that the donor's Roche lobe shrinks in response to the mass transfer. Unstable mass transfer leads to a common envelope (CE), a gas envelope that engulfs both components~\citep[see, e.g.,][for a review]{2013A&ARv..21...59I}. As the two components orbit inside this CE, some of the orbital energy and angular momentum can be transferred to the CE. Sometimes this exchange of orbital energy for CE binding energy leads to a successful CE ejection, leaving behind a binary in a much tighter orbit compared to the pre-CE orbital parameters. \rr{This depends on the CE efficiency $\alpha$ and binding energy $\lambda$, which \textsc{COSMIC} models according to the $\alpha$--$\lambda$ parameterization~\citep[see, e.g.][and references therein]{2002MNRAS.329..897H}; we adopt the default \textsc{COSMIC} values of $\alpha = 1$ (100\% CE efficiency) and the variable $\lambda$ prescription from~\citet{2014A&A...563A..83C}.} The mass ratio of the system at the onset of each episode of mass transfer is key to determining whether the mass transfer is stable or unstable. Here, we use the critical mass ratio $q_\mathrm{crit}$ prescription from~\citet{2014A&A...563A..83C} to determine mass transfer stability. Conversely, the mass transfer, whether stable or unstable, affects the mass ratio of the system, as one star loses mass and may donate some of its mass to its binary companion. We assume the default \textsc{cosmic} prescriptions for the rate of mass loss and mass accretion, including Eddington-limited accretion onto BHs~\citep{2002MNRAS.329..897H}. \rr{The stability of mass transfer and, in the case of unstable mass transfer, the outcome of a CE phase, are major uncertainties in binary population synthesis models, and the prescriptions we use here are only approximate descriptions of a complex multidimensional process. Section~\ref{sec:discussion} discusses how different assumptions about mass transfer stability and the CE may impact our results.}

Another important event in determining the relative masses and orbital parameters of the binary is the supernova at the end of each star's life, giving rise to the NS or BH. The star may lose mass in the supernova explosion, affecting the mass ratio of the system. We use the delayed supernova model in \textsc{cosmic} to determine remnant NS and BH masses~\citep{2012ApJ...749...91F}. 
Supernova mass loss also affects the orbit. Even if the mass loss is spherically symmetric, the binary can be disrupted if enough mass is lost~\citep{1961BAN....15..265B}. If there are asymmetries in the mass loss or anisotropy in the neutrino emission, the supernova may impart a natal kick onto the newborn compact object~\citep{1975Natur.253..698K}. \rr{Depending on the magnitude and orientation of an object's natal kick $v_k$ relative to its pre-explosion orbital velocity $v_0$}, the kick can either widen or shrink the orbital separation and impart eccentricity on the system.
The natal kick imparted by the second supernova can dramatically impact whether or not the compact binary system merges within a Hubble time.
\rr{On average, randomly oriented natal kicks tend to widen the orbital separation and increase the time that the system would take to merge~\citep[e.g.,][]{1999A&A...346...91B}. However, if $v_k < 2 v_0$ and the angle $\cos\theta$ between $v_k$ and $v_0$ exceeds $-v_k/ 2v_0$, conservation of energy and angular momentum will cause the orbit to tighten~\citep[e.g.,][]{1983ApJ...267..322H,1996ApJ...471..352K}. In other words, if a significant fraction of the kick velocity imparted by the second supernova happens to be anti-parallel to the object's pre-explosion velocity in the orbital plane, the binary will tighten. In some cases, this ``lucky" kick might enable an initially wide binary to merge within a Hubble time.}

For the most part, we use the default kick prescriptions in \textsc{cosmic}. For NSs born in core-collapse supernovae, kicks are drawn from a Maxwellian distribution with dispersion parameter $\sigma = 265$ km/s~\citep{HobbsLorimer2005}, while for NSs born in electron-capture supernovae, the dispersion parameter is $\sigma = 20$ km/s~\rr{\citep{2019MNRAS.482.2234G,2020ApJ...898...71B}}. \rr{While there is general consensus that electron-capture supernovae result in small kicks with $\sigma \lesssim 50$ km/s, the exact prescription is uncertain and ranges from $20 < \sigma < 50$ km/s in different models~\citep{2004ApJ...612.1044P,2006ApJ...644.1063D,2008MNRAS.386..553I}}. We take the mass ranges for electron-capture supernovae from~\citet{2004ApJ...612.1044P}. Meanwhile, BHs receive fallback-modulated kicks~\citep{2012ApJ...749...91F}. We also consider a model variation in which BHs receive no kicks. 

\subsection{\textsc{cosmic} setup}
\label{subsec:cosmic}

We model three different metallicities in our \textsc{cosmic} simulations: $Z_\odot$, $0.1\,Z_\odot$ and $0.05\,Z_\odot$. Out of the initial ZAMS population, we track systems that result in BBH and NSBH systems. We distinguish between BBH and NSBH systems that merge within a Hubble time (the ``merging" population) and those that do not merge (the ``non-merging" population). Throughout this work, we define the mass ratio $q = m_2/m_1$, where $m_2$ refers to the initially less massive star (at ZAMS). However, when discussing GW observations, $m_2$ refers to the lighter compact object in the binary (since it is unknown which compact object came from which star). Thus, GW observations always have $q \leq 1$. 

Our parameter settings are mostly the same as the default configuration in \textsc{cosmic} version 3.4.0, with a few changes that we highlight here. The initial masses, eccentricities, separations, and binary fractions are drawn from independent distributions. The initial (ZAMS) primary masses follow the distribution from \cite{2001MNRAS.322..231K} \rr{and we simulate ZAMS masses up to $150\,M_\odot$. At each primary mass, initial mass ratios are drawn from a uniform distribution.} The initial orbital periods and eccentricities follow the distribution from \citet{2012Sci...337..444S}; \rr{i.e., a power law in orbital period (in $\log_{10}$ space) with slope $-0.55$ and a power law in eccentricity with slope $-0.45$.} We use a binary fraction of 0.7. We set solar metallicity $Z_\odot = 0.017$. \textsc{cosmic} uses the ``match" parameter to determine the convergence of each simulated population, ensuring that population statistics are accurate (see~\citealt{2020ApJ...898...71B} for more details). In our simulations, we fix the match parameter to be $-6.0$, which is stricter than the default value of $-5.0$. We adopt the critical mass ratios $q_\mathrm{crit}$ from~\citet{2014A&A...563A..83C} to determine whether mass transfer is stable or unstable during Roche lobe overflow. The maximum and minimum He-star masses that result in electron capture supernovae, which impart relatively small natal kicks on the NSs they produce, are taken from \cite{2004ApJ...612.1044P}. According to our prescription in \textsc{cosmic}, the resulting neutron stars are all born with $\sim1.3\,M_\odot$.


    

\section{Results}
\label{sec:results}
In Section~\ref{sec:formation-channels}, we group merging and non-merging BBHs and NSBHs with similar formation histories and describe their evolution from ZAMS masses to compact object binaries.  We then focus on the pathways leading to asymmetric systems (Section~\ref{sec:asymmetric}) and highlight the impact of natal kicks (Section~\ref{sec:kicks}). 

\subsection{Formation channels}
\label{sec:formation-channels}

\begin{table*}
    \begin{tabular}{c || l | l}
            & Before first supernova & After first supernova \\
        \hline \hline 
        Stable Mass Transfer & \textcolor{yellow}{\bf CEA}: CE After & \textcolor{cyan}{\bf SMT}: Stable Mass Transfer \\
         & \textcolor{cyan}{\bf SMT}: Stable Mass Transfer & \textcolor{red}{\bf CEB}: CE Before \\
        \hline
        Common Envelope (CE) & \textcolor{green}{\bf CEBA}: CE Before and After  & \textcolor{green}{\bf CEBA}: CE Before and After \\
        & \textcolor{red}{\bf CEB}: CE Before & \textcolor{yellow}{\bf CEA}: CE After \\
        & & \textcolor{violet}{\bf CE+MRR}: Mass Ratio Reversal\\
        \hline
    \end{tabular}
    \caption{\rr{Summary of formation channels considered in this work, as described in Section~\ref{sec:formation-channels}.}}
    \label{table:formation-channels}
\end{table*}

\begin{figure*}
    \centering
    \includegraphics[width=15cm]{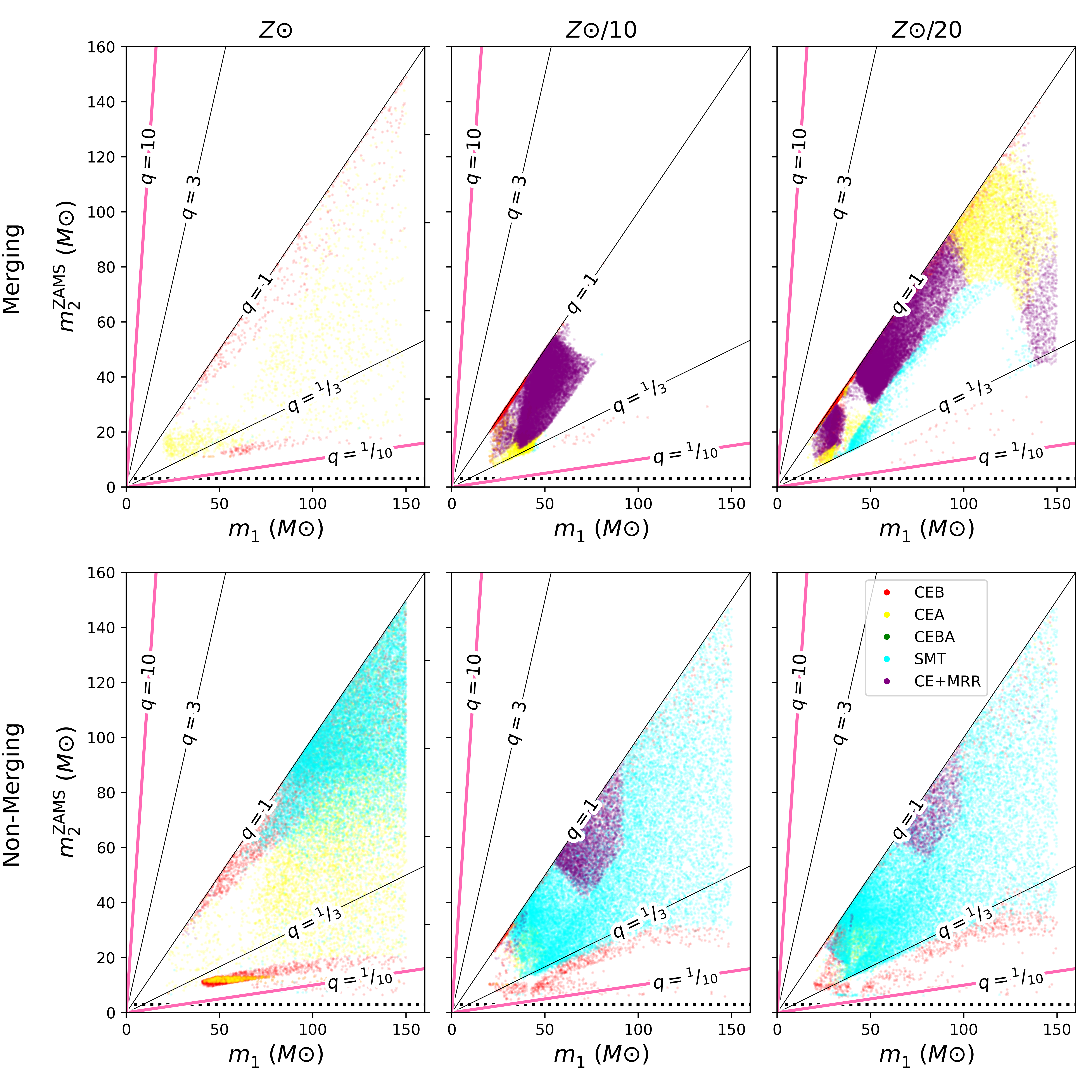}
    \caption{ZAMS masses of merging and non-merging BBH and NSBH systems for models varying the metallicity $Z$. Systems are colored by formation channel. The  horizontal dashed line shows the split between NSBHs and BBHs. The black diagonal lines show mass ratios of 3, 1, and \(\frac{1}{3}\). The pink line runs along the approximate mass ratio of GW190814, \(\frac{1}{10}\) and 10, which are indistinguishable in GW observations.}
    \label{zamsmass}
\end{figure*}


\begin{figure*}
    \centering
    \includegraphics[width=16cm]{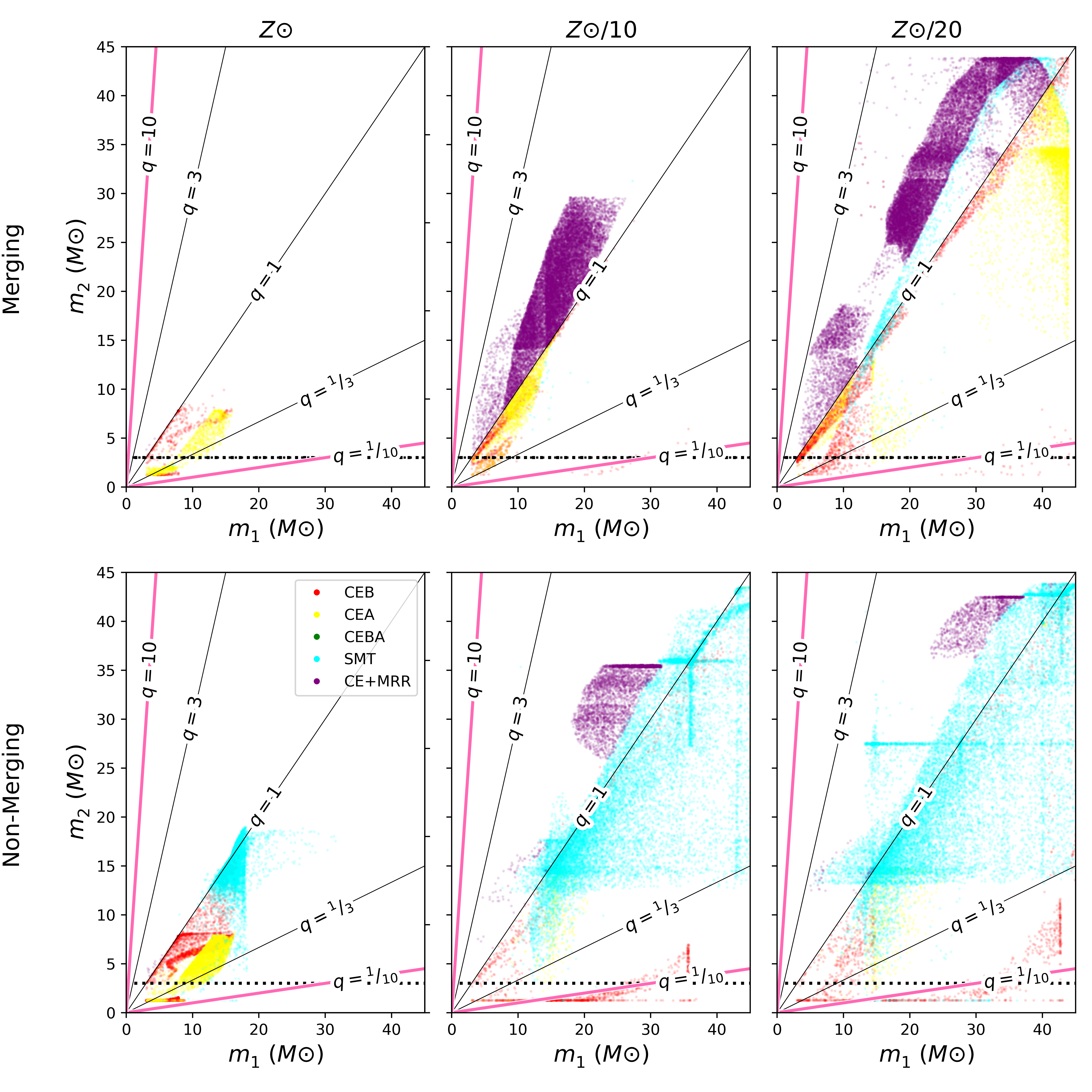}
    \caption{Final component masses of BBH and NSBH systems with the same models as in Figure \ref{zamsmass}, colored by formation channel.}
    \label{finalmass}
\end{figure*}

We split the binary systems into five different formation channels. Each binary experiences two supernovae during its evolution, in which the faster-evolving star collapses to a compact object followed by the collapse slower evolving star. The binary generally undergoes phases of mass transfer both before and after the first supernova.
We characterize systems by whether each phase of mass transfer is unstable, leading to a CE, or stable; \rr{see Table~\ref{table:formation-channels} for a summary of the channels}. The CEB (``Common Envelope Before") channel describes systems that experience a CE before the first supernova but stable mass transfer after the first supernova. CEA (``Common Envelope After") indicates stable mass transfer before the first supernova and a CE phase after the first supernova. Systems with a CE phase after the first supernova that also undergo mass ratio reversal (MRR), in which the initially more massive star ends up as the less massive BH in the binary, follow the CE+MRR channel. 
 CEBA (``Common Envelope Before and After") systems go through CE both before and after the first supernova. Systems with only stable mass transfer (SMT) before and after the first supernova, implying no CE, follow the SMT channel. The SMT and CE+MRR channels are equivalent to Channels A and B, respectively, from~\citet{2020ApJ...899L...1Z}, who identified them as possible evolutionary histories for GW190814.

Figure \ref{zamsmass} shows the ZAMS component masses that form systems in certain channels. Systems with similar ZAMS masses at the same metallicity tend to follow similar evolutionary pathways, leading to clusters in the $m_1^\mathrm{ZAMS}$--$m_2^\mathrm{ZAMS}$ plane. Across all three metallicities and among both merging and non-merging systems, CEB systems (red) form two clusters. One cluster runs along the diagonal $m^\mathrm{ZAMS}_1 = m_2^\mathrm{ZAMS}$ line and the other has $m_1^\mathrm{ZAMS} > 3 m_2^\mathrm{ZAMS}$. To undergo CE before the first supernova, CEB systems generally need to start with a low initial mass ratio ($q < q_\mathrm{crit}$, where $q_\mathrm{crit}$ is the critical mass ratio determining whether mass transfer is unstable or stable) in order to trigger a CE phase. However, systems in the cluster on the diagonal have some of the most symmetric initial mass ratios of the population. According to the \textsc{cosmic} prescriptions, these highly symmetric systems start a CE phase  because both stellar components overflow their Roche lobe radii at the same time during the contact stage. Because the two stars start off with roughly equal masses, they evolve at the same rate, reaching Roche lobe overflow at the same time.

The CEA systems (yellow) in Figure \ref{zamsmass} cover a broad mass range and have mostly intermediate initial mass ratios. At the lowest metallicity we consider ($Z_\odot/20$, rightmost column), systems with higher ZAMS masses are able to merge through the CEA channel. At higher metallicities, systems with high initial total masses still experience the CEA channel, but because they experience stronger winds, they are less likely to merge according to our simulations. We caution that rapid population synthesis simulations like \textsc{cosmic} use prescriptions for stellar expansion and mass loss that \rr{are based on extrapolating lower mass stellar sequences and} may not accurately describe the highest-mass systems. This affects the trends with mass and metallicity that we report here~\citep[e.g.,][]{2022MNRAS.512.5717A,2022arXiv221210924B}. \rr{In particular, recent detailed simulations suggest that BBHs with components up to $\approx30\,M_\odot$ can merge at solar metallicity because their progenitor stars lose their hydrogen envelopes early and therefore avoid significant winds associated the supergiant phase, unlike their lower mass counterparts that expand as red supergiants after the main sequence~\citep{2022arXiv221210924B}.}

The CE+MRR systems (purple) start off roughly symmetric (close to the $q = 1$ diagonal). Figure~\ref{zamsmass} shows a cutoff based on initial total mass that divides merging and non-merging CE+MRR systems. For $Z = 0.1\,Z_\odot$, systems with initial total masses $\gtrsim 110\,M_\odot$ do not merge, while the less massive systems do. 
The less massive systems experience a greater decrease in separation during the CE phase, so they are more likely to merge within a Hubble time.

CEBA systems (green) occur very infrequently for both merging and non-merging systems. The modeling assumptions shown here do not produce any CEBA systems, although we found some CEBA systems in model variations using an ``optimistic" CE prescription. Because it is so rare, we do not discuss the CEBA channel further. 

SMT systems (cyan) cover a wide initial mass range, but rarely merge. In our models, the SMT channel leads to mergers only at lower metallicities (a few at $Z_\odot/10$, and more at $Z_\odot/20$. As we discuss later, this finding is sensitive to uncertainties in modeling mass transfer stability.

Figure \ref{finalmass} shows the final masses (just before coalescence for merging systems and after a Hubble time for non-merging systems) that result from the different formation channels. We find that 72\% of NSBHs across the models shown evolved through the CEB channel (36\% of merging systems and 77\% of non-merging systems), whereas only 6\% of BBHs evolved through the CEB channels (5\% of merging systems and 6\% of non-merging systems). The trend toward higher final compact object masses at lower metallicities results from the stellar wind prescription; at lower metallicities, stars experience much weaker winds and do not lose as much mass. As we noted earlier, however, our prescriptions in \textsc{cosmic} may not accurately describe the evolution of the most massive stars, so these trends with metallicity are not robust. Indeed, stellar evolution calculations show that $30\,M_\odot$ BHs can form at solar metallicity~\citep{2022arXiv221210924B}.  

Another feature that is apparent from comparing the ``Merging" and ``Non-Merging" panels in Figures \ref{zamsmass} and \ref{finalmass} is that very few of the systems that undergo stable mass transfer after the first supernova (the SMT and CEB channels) merge, compared to systems that instead undergo CE after the first supernova (predominantly the CEA and CE+MRR channels). This may be an artifact of our mass transfer stability prescriptions, as recent work has found that rapid population synthesis codes may underestimate the relative contribution of SMT compared to CE to the production of BBH mergers~\citep{2019MNRAS.490.3740N,2021A&A...651A.100O,2021ApJ...922..110G}.




\subsection{Forming asymmetric systems}
\label{sec:asymmetric}

As noted in the previous subsection, the majority of merging BBHs and NSBHs experience CE after the first supernova, when the system consists of a BH and a hydrogen-rich star (Channel CE+MRR or CEA; CEBA systems occur very rarely). The successful ejection of the CE removes orbital energy from the system, shrinking the orbital period sufficiently so that the system can eventually merge in a Hubble time. However, as pointed out by~\citet{2020ApJ...899L...1Z}, this standard CE evolution rarely leads to compact object binaries more asymmetric than 1:5, and almost never to asymmetries greater than 1:10. 

The most asymmetric merging system resulting from CE+MRR (purple points in Figure~\ref{finalmass}) has $q = 3.14$, which is indistinguishable from $q = 0.32$ in GW observations. (Because the mass ratio is calculated as the mass of the initially less massive secondary divided by the mass of the initially more massive primary, CE+MRR compact object binaries have mass ratios greater than 1.) Our findings agree with~\citet{2022ApJ...933...86Z} who find that MRR rarely leads to mass ratios smaller than $1/3$.

The ``CEA" points (yellow) in Figure~\ref{finalmass} show that this channel also rarely produces systems with extreme mass ratios. In our models, the most asymmetric merging CEA system has $q = 0.16$, while the most asymmetric non-merging CEA system has $q = 0.13$. Across all three metallicities, these extreme mass ratio CEA systems started with low ZAMS masses, forming clusters in the lower left corner of each subplot in Figure \ref{zamsmass}. When these systems go through stable mass transfer before the first supernova, the initially more massive primary transfers enough mass to the secondary, so that the secondary ends up being more massive than its companion. However, the mass ratio does not reach the asymmetry needed for a CE phase before the first supernova. The primary, now less-massive star, then loses mass during its supernova, increasing the asymmetry. At this point, the two masses are different enough to trigger a CE phase, which then shrinks the orbit and distributes the mass to be roughly even between the two components again. The only way to then end up with an asymmetric compact object binary is if the second-born compact object loses enough mass during its supernova explosion. In some rare cases, this second supernova can lower the mass ratio to $q < 0.2$. Because only low mass systems experience significant supernova mass loss according to our prescription, these systems all end with low masses and are generally NSBHs rather than BBHs. The difference between the component masses of even the most asymmetric CEA systems is only a few solar masses.

We find that the majority of asymmetric systems come from the CEB channel (red points in Figure~\ref{finalmass}), and these systems are the only ones that can produce mergers with $q \simeq 0.1$. Both Figure \ref{zamsmass} and Figure \ref{finalmass} show two main groups of CEB systems in each panel. Systems that started with relatively low initial $q$ in Figure \ref{zamsmass} were the ones that ended up low-$q$ in Figure \ref{finalmass}. Because these systems start with extreme mass ratios, they go through CE before the first supernova. However, as we explain in the following section, because there is no second CE phase after the supernova, the systems rely on kicks to tighten the orbit and merge.

\subsection{Impact of natal kicks}
\label{sec:kicks}

\begin{figure*}
    \centering
    \includegraphics[width=16cm]{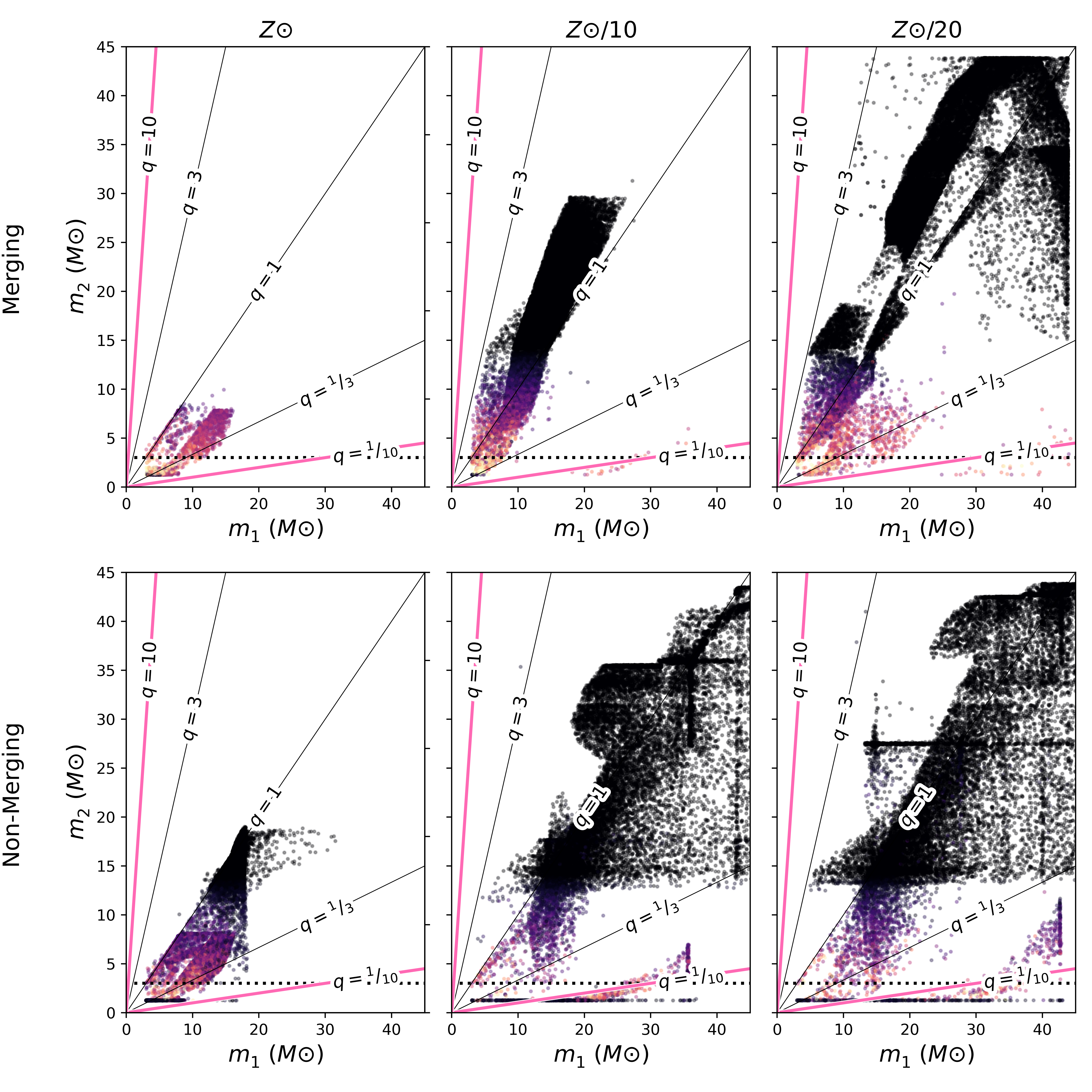}
    \caption{Component masses of merging and non-merging NSBHs and BBHs with the same models as in Figure \ref{zamsmass}. Systems are colored by the natal kick magnitude imparted by the second supernova.}
    \label{kicks}
\end{figure*}

\begin{figure*}
    \centering
    \includegraphics[width=16cm]{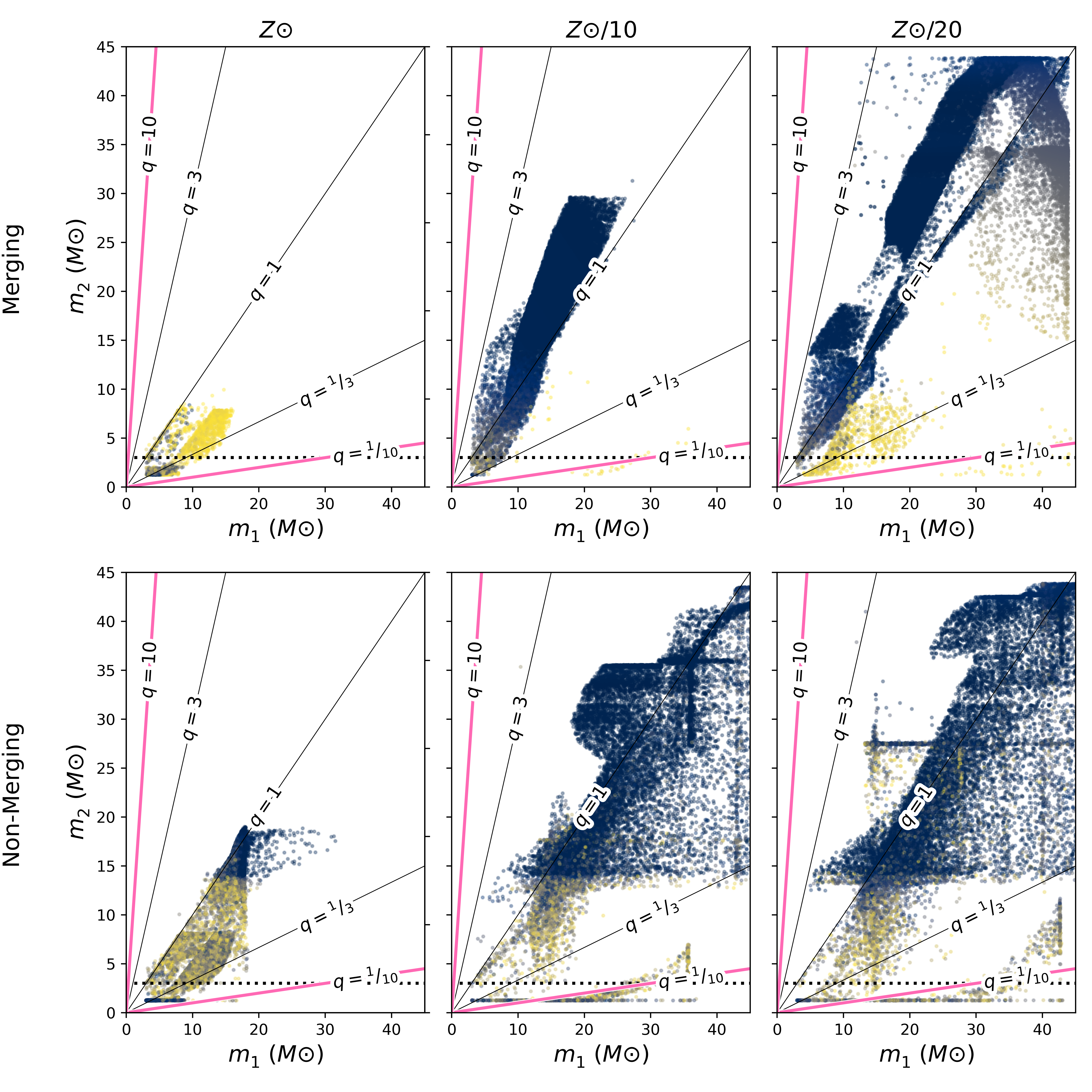}
    \caption{Component masses of merging and non-merging NSBH and BBH systems with the same models as in Figure \ref{zamsmass}, colored by eccentricity.}
    \label{ecc}
\end{figure*}

\begin{figure*}
    \centering
    \includegraphics[width=16cm]{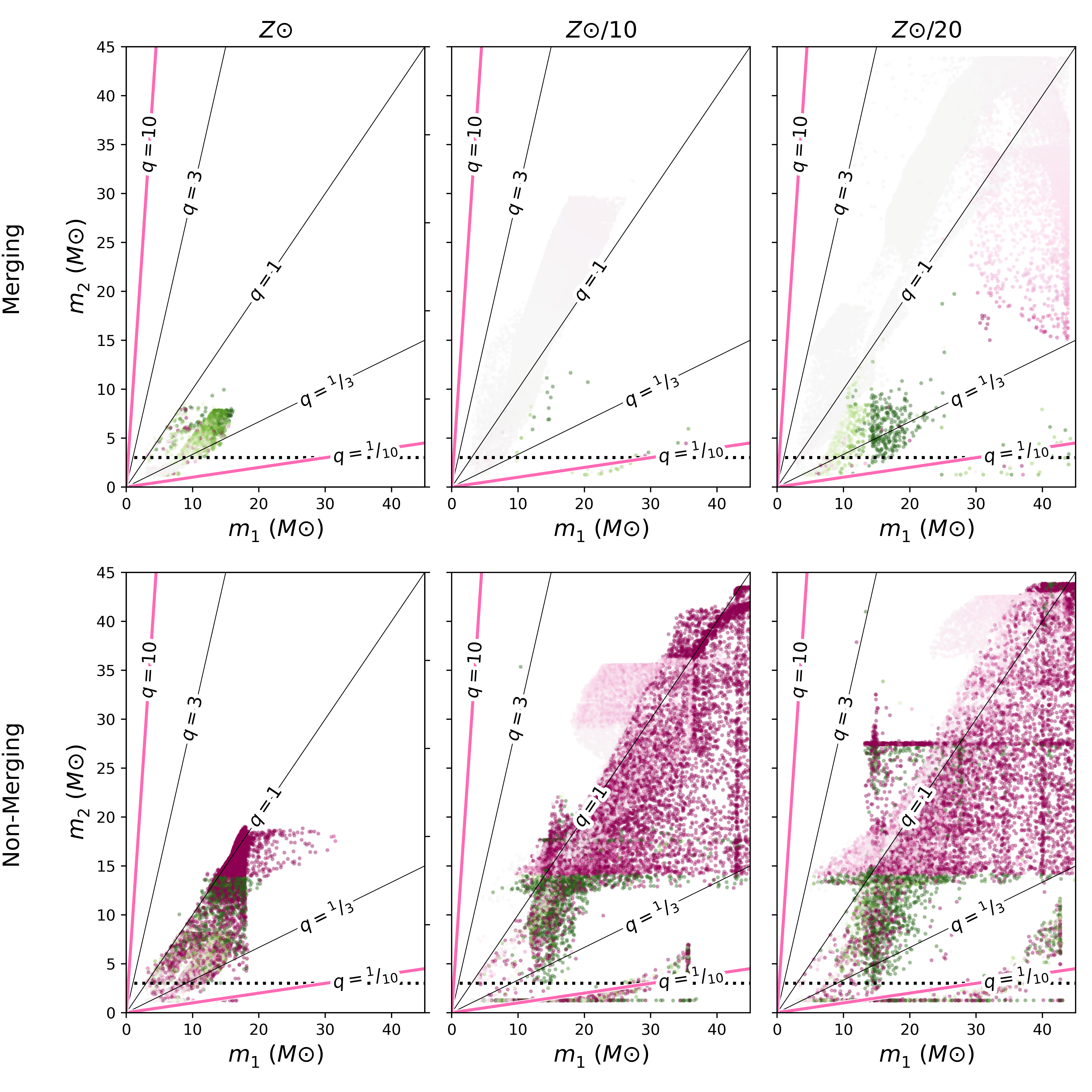}
    \caption{Component masses of merging and non-merging NSBH and BBH systems with the same models as in Figure \ref{zamsmass}. Systems are colored by the difference in orbital separation before and after the secondary receives its natal kick. Green indicates that the components came closer together after the kick, while magenta indicates that moved farther away. White points showed little to no change in separation.}
    \label{sep}
\end{figure*}

The importance of natal kicks for merging the asymmetric CEB systems is evident in Figure \ref{kicks}, which shows the difference in second kick magnitudes received by merging systems (top panel) compared to those received by non-merging systems (bottom panel). The role of natal kicks in merging binary NSs was recently highlighted by \citet{2022arXiv221115693G}. In general, natal kicks tend to be smaller for larger $m_2$, a consequence of the assumed supernova prescription in which smaller compact objects are born from more explosive supernovae that impart larger kicks~\citep{2012ApJ...749...91F}. Thus, NSs generally receive larger kicks than BHs, meaning the second kick for NSBHs is typically larger than the second kick for BBHs. The exception to this is the line of dark-colored non-merging points with $m_2 = 1.3\,M_\odot$, which result from electron capture supernovae. Electron capture supernovae are thought to impart very small kicks and produce neutron stars within a narrow range of masses~\citep{2010ApJ...719..722S,2018ApJ...865...61G}.  

Focusing on the asymmetric systems, it is clear that the most asymmetric merging systems with $q < 0.1$ generally receive much stronger kicks ($\sim250$ km/s) than the non-merging systems ($\lesssim 100$ km/s). 
The link between natal kicks and merger rates is further demonstrated in Figure \ref{ecc} which shows the eccentricity immediately after that second kick for merging and non-merging systems. Every merging asymmetric system has an eccentricity close to 1 after the kick, while non-merging systems have, on average, far less eccentric orbits after the kick. At the same orbital separation, eccentric systems merge in less time compared to perfectly circular orbits~\citep{1964PhRv..136.1224P}. 

However, as indicated by the bottom panels of Figure \ref{kicks} and Figure \ref{ecc}, there are many systems that do not merge within a Hubble time even after receiving a relatively strong kick. We find that these systems typically are not kicked in the right direction to shrink the orbit and are instead pushed apart. Figure \ref{sep} shows the impact of the second kick on the orbital separation of merging and non-merging systems. Green-colored points indicate that the orbit shrank after the kick, and magenta means that the components were pushed farther apart. Almost all merging asymmetric systems in this figure are brought closer together by the kick, while most of the non-merging systems are either pushed apart or do not show much change. The asymmetric systems that did merge just happened to receive a kick that had a large enough magnitude and also happened to be in the right direction.

To further test the role of natal kicks, we also ran a set of simulations under the assumption that BH kicks are always zero (keeping all other parameters the same). In these populations, asymmetric BBH mergers are even more rare compared to the case where BH kicks are nonzero. Without BH kicks, no BBH mergers with $q < 0.1$ formed out of a total of 54,847 BBH mergers. However, with BH kicks, five out of 48,782 BBH mergers had $q < 0.1$. \rr{While this difference is small compared to other population synthesis uncertainties (see Section~\ref{sec:discussion}), it further confirms that natal kicks play a role in causing asymmetric systems to merge.}






\section{Discussion}
\label{sec:discussion}

Traditionally it was thought that in isolated binary evolution, the progenitors of most merging compact binary systems underwent unstable mass transfer between the first-born compact object and its companion hydrogen-rich star, leading to a common envelope (Channel CEA or CE+MRR in this work); see, e.g. the review by~\citep{2022PhR...955....1M}. A common envelope can exchange its binding energy for orbital energy, allowing for the orbit to tighten after the common envelope is ejected and setting the binary up to successfully merge in a Hubble time. Indeed, according to our \textsc{cosmic} runs, the vast majority of merging systems undergo common envelope after the first supernova. However, a common envelope can only occur if the mass of this first-born compact object is sufficiently smaller than the mass of its hydrogen-rich companion. When this is the case, after the hydrogen-rich star continues to evolve and lose a lot of mass, it ends its life as a compact object with a similar mass to the first-born compact object in the binary; in other words, a symmetric system. The second-born compact object can even be more massive, despite starting off as the smaller star (the CE+MRR case), but we find that it very rarely exceeds the mass of the first-born compact object by more than a factor of 3. Therefore, we find that in order to end up with a highly asymmetric compact object binary, the progenitor to the second-born compact object must be sufficiently low mass before its supernova, putting it at a similar mass as the first-born compact object. This means that the mass transfer at the compact object + H-rich star stage is stable, as in the CEB case, and there is no CE ejection to rapidly tighten the orbit. 
Instead, a similar tightening of the orbit can be achieved if the second-born compact object receives a natal kick in the right direction. In our set of simulations, we find that the only way for highly asymmetric systems to merge is if they receive such a lucky natal kick. 

An intriguing consequence of our results is the prediction that asymmetric mergers are more common among GW sources involving a low mass component, including NSBH and BBH with small $m_2$, because smaller compact objects generally receive higher natal kicks. Interestingly, this is consistent with the emerging picture from GW observations. There are hints that mass ratios $q < 1$ are more common among small $m_2$ compared to large $m_2$, with the pairing probability more strongly preferring symmetric binaries when $m_2 > 5\,M_\odot$~\citep{2022ApJ...931..108F,2022ApJ...933L..14L}. This will be further tested with more precise measurements of the merging compact binary mass distribution from future GW observations. Additionally, the spins, particularly spin-orbit (mis)alignments, of GW events will help determine supernova natal kicks.

There are several caveats to this work. Our conclusions rely on understanding which systems experience common envelope versus stable mass transfer. Recently, the mass transfer stability prescriptions typically used in rapid population synthesis like \textsc{cosmic} have come into question~\citep{2019MNRAS.490.3740N,2021A&A...651A.100O,2021ApJ...922..110G,2022ApJ...931...17V,2022MNRAS.513.4802A}. \rr{In particular, after the formation of the first compact object in the binary, mass transfer between the H-rich donor and the compact object accretor strongly impacts whether or not the binary can merge in a Hubble time. The systems that end up as asymmetric compact object binaries tend to be fairly symmetric at the onset of this mass transfer phase. 
According to our assumed critical mass ratio thresholds from~\citet{2014A&A...563A..83C}, this phase of mass transfer tends to be stable for such mass ratios (CEB or SMT channel). If some of these systems instead experienced unstable mass transfer (i.e. if $q_\mathrm{crit}$ was closer to unity), more of them would undergo common envelope (e.g. the CEBA rather than the CEB channel or CEA rather than SMT). We would likely predict a higher number of asymmetric mergers overall because some of them would be assisted by a common envelope rather than relying only on a natal kick to tighten the orbit. Nevertheless, we do not expect different $q_\mathrm{crit}$ assumptions to affect our qualitative conclusions about the formation channels that result in asymmetric compact object binaries.} 

\rr{In addition to the $q_\mathrm{crit}$ prescription that determines whether the donor initiates a common envelope phase, the outcome of the common envelope is a key uncertainty in our calculations. We have modeled the common envelope with the commonly used $\alpha$--$\lambda$ energy prescription with $\alpha = 1$. The efficiency $\alpha$ is uncertain, with higher (lower) values of $\alpha$ generally predicting larger (smaller) orbital separations post-common envelope. Furthermore, as highlighted recently by detailed stellar modeling, the $\alpha$--$\lambda$ prescription that we adopt here is a simplification~\citep{2021A&A...645A..54K,2022MNRAS.511.2326V,DiStefano_Kruckow_Gao_Neunteufel_Kobayashi_2022,Hirai_Mandel_2022,Roepke_De,Yarza_Everson_RamirezRuiz_2022,Wilson_Nordhaus_2022,Renzo_Zapartas_Justham_Breivik_Lau_Farmer_Cantiello_Metzger_2023}. According to our simulations, asymmetric compact object binaries usually experience a common envelope before the first supernova, but not after the first supernova (the CEB channel). For these binaries, the common envelope is not important in shrinking the orbit and resulting in a compact object merger. Nevertheless, common envelope uncertainties may impact our conclusions if we are significantly misestimating the number of binaries that can survive the initial common envelope phase.}

There are additional uncertainties in binary population synthesis that we did not explore in this work. For example, it has been proposed that super-Eddington accretion onto the first-born BH can dramatically increase its mass and lead to more asymmetric systems~\citep{2021A&A...647A.153B,2022ApJ...933...86Z,2022arXiv220613842B}. However, such conservative mass transfer is not as efficient as non-conservative mass transfer in shrinking the orbit, and so this scenario may not contribute much to the merger rate. Significant accretion would also generally spin up the BH~\citep[although see, e.g.][]{2023arXiv230201351L}, while the spin of the primary BH in GW190814 is tightly constrained to be near zero. 

Another uncertainty that should be explored further in future work is the supernova prescription, which sets the remnant mass and the natal kick. Recent studies have explored the impact on updated remnant mass and natal kick prescriptions on various observable NS and BH populations, showing that they can significantly affect GW merger rates and masses as well as the properties of Galactic double NSs and BHs in X-ray binaries~\citep{2016MNRAS.461.3747B,2018ApJ...866..151A,2018MNRAS.480.5657B,2020ApJ...891..141G, 2021MNRAS.500.1380M,2022MNRAS.516.2252O}. Meanwhile, \citet{2022A&A...657L...6A} argued that explodability fluctuations in core-collapse supernovae may create systems like GW190814 because the second supernova may shed significantly more mass, leaving behind a much smaller compact object, than we assume in the standard prescription. If this is the case, then the CEA channel could produce more asymmetric systems. Supernova uncertainties also extend to the uncertain electron-capture supernova prescription, which may affect our results because it determines how many NSs experience significant kicks.

\rr{Stellar wind mass loss is another key process that affects the remnant masses. As we mentioned in Section~\ref{sec:formation-channels}, our prescriptions for stellar expansion and wind mass loss are uncertain, particularly in their dependence on metallicity. Although these uncertainties are unlikely to affect our main conclusions, we expect the trends we find with metallicity to be less robust.}

It is also important to note that this work considers only the isolated binary evolution channel for forming NSBH and BBH. Other formation channels, such as evolution from stellar triples~\citep{2022ApJ...937...78M} and dynamical assembly in dense stellar environments~\citep{2020ApJ...901L..34Y} may contribute to, or in fact dominate, the rate of asymmetric mergers.

\section{Conclusion}
\label{sec:conclusion}
In this work, we simulated large populations of binary stars using the \textsc{cosmic} rapid population synthesis code. We explored asymmetric BBH and NSBH systems with $q \lesssim 0.2$ with a particular focus on the most asymmetric systems with $q < 0.1$, and studied how these systems form and merge. The main results of this work are as follows:
\begin{itemize}
\item Most NSBH or BBH binaries with extreme mass ratios start with low initial mass ratios $q^\mathrm{ZAMS} < 0.3$ and undergo a CE phase before the first supernova.
\item After the collapse of the first object, the mass ratio is close to unity, causing the second phase of mass transfer to be stable. This stable mass transfer fails to shrink the orbit sufficiently.
\item In order for systems with $q \lesssim 0.1$ to merge, they appear to require a kick of $\sim250$ km/s to shrink the orbit enough.
\item The differences between NS and BH kicks affect the differences between the rates of asymmetric NSBH and BBH mergers.
\end{itemize}

\acknowledgments We thank Katie Breivik, Camille Liotine and Mike Zevin for their help with \textsc{cosmic}. MF was partially supported by NASA through NASA Hubble Fellowship grant HST-HF2-51455.001-A awarded by the Space Telescope Science Institute, which is operated by the Association of Universities for Research in Astronomy, Incorporated, under NASA contract NAS5-26555. 
CK is supported by the Riedel Family Fellowship.
VK is grateful for support from a Guggenheim Fellowship, from CIFAR as a Senior Fellow, and from Northwestern University, including the Daniel I. Linzer Distinguished University Professorship fund. 
This work utilized the computing resources at CIERA provided by the Quest high performance computing facility at Northwestern University,
which is jointly supported by the Office of the Provost,
the Office for Research, and Northwestern University Information Technology, and used computing resources at
CIERA funded by NSF PHY-1726951.
This material is based upon work supported by NSF's LIGO Laboratory which is a major facility fully funded by the National Science Foundation.

\bibliographystyle{aasjournal}
\bibliography{references}








\end{document}